\documentclass[a4paper]{nature}
\usepackage[utf8]{inputenc}
\usepackage{amsmath}
\usepackage{amsfonts}
\usepackage{amssymb}
\usepackage{graphicx}
\usepackage{parskip}
\usepackage{siunitx}
\usepackage{color}
\usepackage[justification=centering,labelfont=bf]{caption}
\usepackage[capitalize]{cleveref}
\usepackage[T1]{fontenc}
\usepackage{authblk}
\usepackage{etoolbox}
\usepackage{bibunits}

\usepackage[normalem]{ulem}
\usepackage{cancel}

\DeclareCaptionLabelFormat{adja-page}{\hrulefill\\#1 #2 \emph{(previous page)}}

\makeatletter
\let\saved@includegraphics\includegraphics
\AtBeginDocument{\let\includegraphics\saved@includegraphics}
\renewenvironment*{figure}{\@float{figure}}{\end@float}
\makeatother
\newcommand{\hematite}{$\alpha$-Fe$_2$O$_3$}

\author{F. P. Chmiel$^{1}$, N. Waterfield Price$^1$, R. D. Johnson$^1$,  A. D. Lamirand$^2$,  J. Schad$^3$, G. van der Laan$^2$, D. T. Harris$^3$, J. Irwin$^4$, M. S. Rzchowski$^4$, C.-B. Eom$^3$  \& P. G. Radaelli$^{1*}$}

\begin{document}
\begin{bibunit}[naturejournals]
\title{Observation of magnetic vortex pairs at room temperature in a planar \hematite{}/Co heterostructure.}

\maketitle

\begin{affiliations}
 \item  Clarendon Laboratory, Department of Physics, University of Oxford, Parks Road, Oxford OX1 3PU, United Kingdom
 \item Diamond Light Source, Harwell Science and Innovation Campus, Didcot OX11 0DE, United Kingdom
 \item Department of Materials Science and Engineering, University of Wisconsin-Madison, Madison, Wisconsin 53706, USA
 \item Department of Physics, University of Wisconsin-Madison, Madison, Wisconsin 53706, USA
\end{affiliations}

\begin{abstract}

Vortices are among the simplest topological structures, and occur whenever a flow field `whirls' around a one-dimensional core.  They are ubiquitous to many branches of physics, from fluid dynamics to superconductivity and superfluidity,\cite{Zurek1985} and are even predicted by some unified theories of particle interactions, where they might explain some of the largest-scale structures seen in today's Universe.\cite{Kibble1976}  In the crystalline state, vortex formation is rare, since it is generally hampered by long-range interactions: in ferroic materials (ferromagnetic and ferroelectric), vortices are only observed when the effects of the dipole-dipole interaction is modified by confinement at the nanoscale,\cite{zheng_2017, yadav_2016, cowburn_1999} or when the parameter associated with the vorticity does not couple directly with strain.\cite{choi_2010}  Here, we present the discovery of a novel form of vortices in \emph{antiferromagnetic} (AFM) hematite (\hematite{}) epitaxial films, in which the primary whirling parameter is the staggered magnetisation.  Remarkably, ferromagnetic (FM) topological objects with the same vorticity and winding number of the \hematite{} vortices are imprinted onto an ultra-thin Co ferromagnetic over-layer by interfacial exchange.   Our data suggest that the ferromagnetic vortices may be \emph{merons} (half-skyrmions, carrying an out-of-plane core magnetisation), and indicate that the vortex/meron pairs can be manipulated by the application of an in-plane magnetic field, H$_{\parallel}$, giving rise to large-scale vortex-antivortex annihilation.  \end{abstract}

\begin{figure}[!h]
\centering
\includegraphics{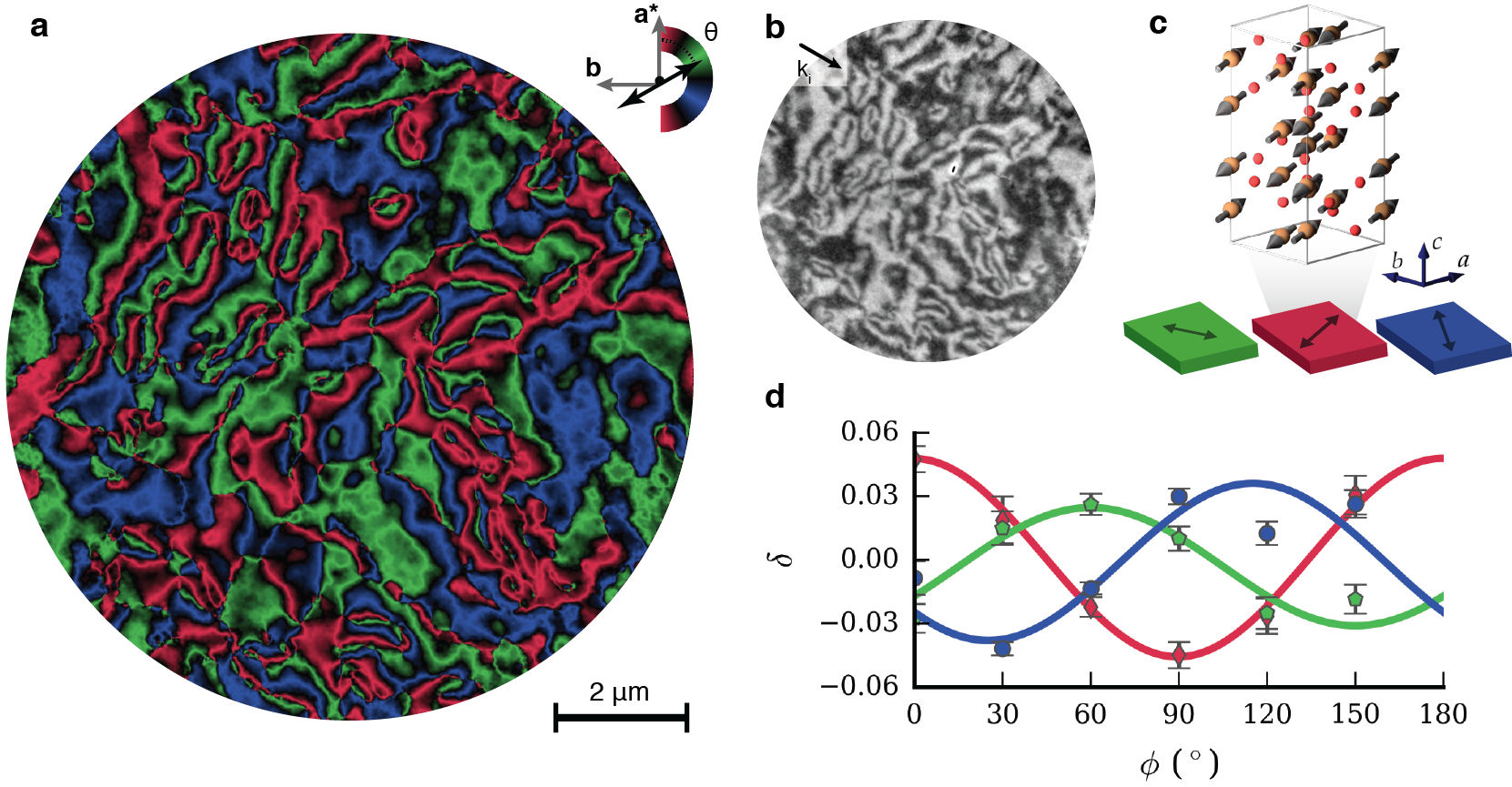}
\caption{\textbf{Vector map of \hematite{} antiferromagnetic domain configuration from x-ray photoemission microscopy. a,} \hematite{} domain structure constructed using the anisotropy of the XMLD signal. \textbf{b,} Measured XMLD signal for a single incident x-ray direction (azimuth), denoted by the solid black arrow. \textbf{c,} Graphic displaying the magnetic unit cell of \hematite{} and the three antiferromagnetic orientation domains. \textbf{d,} Mean dichroism (XMLD signal), as a function of azimuth, for three 15 x 15 pixel regions, one in each of the three identified domains. Solid lines are the fits assuming a single AFM spin direction in each region.}
\label{fig:experiment_setup}
\end{figure}

\hematite{} is a `classic' room-temperature antiferromagnet, with a N\'eel temperature of 948 K and the trigonal corundum crystal structure (space group $R\bar{3}c$). The Fe moments lie in the basal $ab$ plane and are antiferromagnetically ordered along $c$ (\cref{fig:experiment_setup}). We grew 10 nm-thick \hematite{}  epitaxial films with a  Co (1 \si{\nano\metre}) / Al (1 \si{\nano\metre})  overlayer on a (0001)-Al$_2$O$_3$ substrate by magnetron sputtering (see Methods).  The Co layer was confirmed to be ferromagnetically ordered, with  easy-plane anisotropy, by Magneto-Optical Kerr Effect magnetometry (MOKE, see Methods) and its coercive field was found to be $\sim$ 8 \si{\milli\tesla}. Vector-mapped x-ray magnetic linear (for \hematite{}) and circular (for Co) dichroism photoemission electron microscopy (XMLD-PEEM and XMCD-PEEM) images were collected at room temperature and zero applied magnetic field at the Fe-L$_{3}$ and Co- L$_{3}$ edges respectively,  using the I06 beamline at the Diamond Light Source, Didcot,~UK. Exploiting the anisotropy of the XMLD / XMCD signal, data collected at different sample rotations were combined, to determine  absolutely the  direction parallel/anti-parallel to the AFM spins and the Co spin direction at the same positions on the surface of the sample (\cref{fig:experiment_setup}).\cite{hallsteinsen_2017, price_2016, alders_1995, scholl_2002}  We found that the \hematite{} spins are perpendicular to both the crystallographic $c$-axis and to one of the three 2-fold axes (magnetic space group $C2/c$), as previously reported.\cite{chen_2012, marmeggi_1977} The three possible orientations of the AFM moments, related by the broken 3-fold rotation axis of the crystallographic structure,  form an intricate domain structure (\cref{fig:fe2o3_domain_maps} a) at the scale of a few tens of \si{\nano\metre}. One prominent feature of the domain structure is the presence of `pinch-points' between \hematite{} domains of the same orientation. Around these points, domains are typically arranged in a 6-sector pinwheel, colors alternating as either R-G-B-R-G-B or R-B-G-R-B-G  clockwise around their common core (here, `R' represents \hematite{} domains with the AFM spin direction parallel to [2,1,0], `B' to [1,2,0] and `G' to [-2,1,0] in the hexagonal setting of the unit cell, see \cref{fig:experiment_setup} c).  Overall, the domain morphology is strikingly reminiscent both of the disclination domains in liquid crystals\cite{Nehring1972} and of the cloverleaf \emph{ferroelectric} domain patterns imaged  in the hexagonal manganite YMnO$_3$ by piezoelectric force microscopy (PFM), pointing towards a common phenomenology in completely different contexts.\cite{choi_2010, chae_2012}   Remarkably, at the phenomenological level there is an almost complete analogy between AFM ordering in \hematite{} and structural ordering in YMnO$_3$, so that the terms in the Landau free energy expansion that contain only the principal order parameter are exactly the same up to at least the sixth order (See Methods for further details).\cite{Artyukhin2013}   

It is important to emphasise the crucial relationship between the phenomenological Landau free energy and the observation of topological structures such as vortices: in all previously known cases, vortex formation is associated with a phase transition between a phase that is \emph{topologically trivial} (non-superconducting, non-superfluid, disordered liquid crystals etc) and one that is topologically rich enough to support vortices (superconducting, superfluid, nematic or smectic liquid crystals etc.). \cite{Kibble1976}  Therefore, the phenomenological equivalence between YMnO$_3$ and \hematite{} proves that the latter can support vortices in the antiferromagnetic phase, in spite of the differences between the two systems (for example, in the case of \hematite{}, the order parameter is magnetic rather than structural).    More broadly, the Landau free energies of both YMnO$_3$ and \hematite{} approach the U(1) symmetry of the Kibble model in the limit of a small order parameter, which is one way of satisfying the criterion of sufficient topological richness (see Methods).   This being the case, vortices \emph{will form by statistical necessity} when crossing the phase transition as a function of temperature.  At lower temperatures, new vortex/antivortex pairs may form or, conversely, existing vortex/antivortex pairs may annihilate, with the second process being dominant at lower temperatures.  In the case of a thermal transition, the Kibble-Zurek model predicts that the number of vortices/antivortices surviving at low temperatures depends on the quenching rate through the transition, and this relation is also observed in the hexagonal manganites.\cite{meier_2017}  Since our \hematite{} films were grown far below the N\`eel temperature, one would expect that a significant population of vortices and antivortices, both free and bound,  will be frozen in at growth.

On this basis, the Fe L$_{3}$ XMLD-PEEM  images can be straightforwardly interpreted:  the `pinch-points' are the \textit{loci} of 180$^\circ$ domain boundaries. where the AFM order parameter has the same orientation but changes sign across the pinched boundary.  Such boundaries are unfavourable energetically and are reduced to points (lines parallel to the film normal in 3 dimensions) during the domain formation, whilst the most favourable 60$^{\circ}$ domain walls end up dominating. While the XMLD-PEEM method is insensitive to this change in sign of the AFM order parameter, the pinched domain structure allows us to conclude, in analogy to the hexagonal manganites,\cite{choi_2010} that the `pinwheels' we observe are AFM vortices (R-G*-B-R*-G-B*; winding number +1) and anti-vortices (R-B*-G-R*-B-G*; winding number -1), asterisks denoting time inversion. The vortices /anti-vortices form a relatively dense network, with approximately 40 instances over a 80 \si{\micro\metre\squared} area. There are several closely-bound vortex/antivortex pairs, as well as isolated vortices, suggesting that the sample has been quenched without undergoing a Kosterlitz-Thouless transition.\cite{Kosterlitz1973}  Thus far, AFM vortices had only been created in nanoscale disk heterostructures by imprinting from FM vortices,\cite{sort_2006,Wu2011} and have never been previously observed in a homogeneous system. 

Co- L$_{3}$ XMCD-PEEM images (\cref{fig:fe2o3_domain_maps} b) revealed  that the Co film, which at this thickness is magnetically very soft, was found to be strongly textured at the sub-micron scale. The Co spins tended to align strongly \emph{parallel} to the \hematite{} spin direction, so that the \hematite{} and the Co vector maps share a similar morphology (\cref{fig:fe2o3_domain_maps} a and b).  Strikingly, the topological features of  the \hematite{} magnetism are \emph{also} imprinted on the Co over-layer.  In particular, FM vortices and anti-vortices are clearly visible on top of corresponding \hematite{} features with the \emph{same} winding number (\cref{fig:fe2o3_domain_maps} a and b) with the vortex cores being precisely aligned within the resolution of our measurements.  \Cref{fig:fe2o3_co_comparison} displays numerical differentiation of the XMCD-PEEM images, which enabled us to pinpoint exactly the location and nature of the FM vortices and antivortices: clock-wise and anticlock-wise vortices are most apparent in the maps of the curl of the Co magnetisation, while antivortices display an angular alternation of curl in different sectors.  The typical vortex core size, extracted from the full width at half maximum of maps of the magnitude of the Co magnetisation (\cref{fig:fe2o3_co_comparison} c), is of the order of 100 \si{\nano\metre} and appears to be limited by the resolution of our experiment.  Once again, we emphasise that FM vortices were only previously observed in nanodots.\cite{cowburn_1999,wintz_2013} 

In magnetic vortex structures, even with  strong planar-anisotropy, the magnetisation often points out-of-the plane at the vortex core to satisfy  unfavourable exchange and dipole interactions.\cite{shinjo_2000} Indeed, in our XMCD-PEEM images we observe a reduction of the magnitude of the XMCD, sensitive only to the in-plane magnetisation (\cref{fig:fe2o3_co_comparison} c), which is compatible with the presence of an out-of-plane component of the Co spins at the vortex core. This raises the intriguing possibility that the FM topological structures we observe are \emph{merons/antimerons}  (also known as half-skyrmions)\cite{senthil_2004}  rather than planar vortices/antivortices (\cref{fig:fe2o3_co_comparison} a), imprinted by interfacial exchange with  the adjacent AFM vortex/anti-vortex in the \hematite{} film. To test the feasibility of this scenario we  have performed a set of micromagnetic simulations (see Methods for details) where a  fixed, discrete \hematite{} vortex is adjacent to an initially randomized Co layer. We find that the Co layer relaxes into a vortex state with the same topology of the AFM. Notably, at the core of the FM vortex an out-of-plane moment forms, confirming that it is indeed possible to imprint FM merons from an adjacent AFM. Although the out-of-plane magnetic moment of the meron core is not a topological invariant, reversing it by a global rotation in spin space requires `unwinding' the meron, which is strongly disfavoured by the easy-plane anisotropy, meaning the moment at the core should be stable to small external perturbation. 

We found it is possible to manipulate the vortex/meron pairs by the application of moderate magnetic fields. To achieve this, we applied an \textit{ex-situ} 100 \si{\milli\tesla} field parallel to [1,1,0] and re-aligned the sample in the same position under the electron microscope. \Cref{fig:fe2o3_domain_maps,fig:merons} show the remarkable end result of this process: domains not involved in `pinch-points' are largely invariant, (e.g., \cref{fig:merons} c and d), meaning that the large-scale features of the vector map are mostly unchanged. However, large scale vortex/antivortex annihilation has clearly occurred, removing all `pinch-points' from the \hematite{} film, \cref{fig:merons} a and b. This phenomenon is confirmed by the Co L$_{3}$ XMCD-PEEM vector map (\cref{fig:fe2o3_domain_maps} d) of the Co over-layer:  although the Co spins maintain a strong tendency to align along the direction of the \hematite{} staggered magnetisation, the vast majority of FM vortex/merons have disappeared.
 
Our observation of a dense population of coupled AFM/FM vortex pairs in a planar homogeneous \hematite{}/Co heterostructure indicates multiple avenues for exploitation in both fundamental and applied research.  At the fundamental level, this system represents the first purely magnetic realisation of the Kibble-Zurek model with discrete Z$_6$ symmetry, which has been previously studied in as a  structural realisation in YMnO$_3$.\cite{griffin_2012} The close proximity to room temperature of the first-order Morin transition ($\sim$ 260 K for bulk \hematite{}, tuneable by film thickness and doping),\cite{shimomura_2015} below which  the Fe spins flop along the $c$ axis and all AFM vortices must disappear, provides a further opportunity to study the first-order analogue of the Kibble-Zurek model, in which the string density is controlled by nucleation rather than fluctuations.  On the applied side, FM merons can be thought of as topologically protected spin `bits', and could be very appealing for information storage in both pinned and free forms.  A regular array of pinned merons, arranged for example in a cross-point architecture,\cite{yu_2011} could store information in the magnetisaton of the FM meron core, in analogy to proposed memories based on arrays of FM nanodots.\cite{nakano_2011} If a population imbalance between merons and antimerons could be created, free merons could be produced by cooling below the Morin transition. Such a hypothetical device could serve as a source for `meron racetrack memories', similar to those currently considered for skyrmions.\cite{parkin_2008,tomasello_2014}  

\section*{Acknowledgements}

We acknowledge Diamond Light Source for time on Beam Line I06 under Proposal SI16338 and SI15088. We thank S. Parameswaran for discussions and T. Hesjedal and S. Zhang for assistance with initial film growth. The work done at the University of Oxford (F. P. C, N. W. P., R. D. J., and P. G. R.) is funded by EPSRC Grant No. EP/M020517/1, entitled Oxford Quantum Materials Platform Grant. The work at University of Wisconsin-Madison (J.S., J. I., M. S. R. and C.-B. E.) supported by the Army Research Office through Grant No. W911NF-13-1-0486 and W911NF-17-1-0462. R. D. J. acknowledges support from a Royal Society University Research Fellowship. 

\section*{Author Contributions}
F. P. C., N. W. P., R. D. J. and A. D. L. performed the experiment. F. P. C. and A. D. L. performed the data reduction. F. P. C and N. W. P. performed the data analysis. J. S grew the films. D. T. H. made the \hematite{} sputtering target. J. S and F. P. C. characterised the epitaxial relation of the films. J. I. performed the MOKE measurement. G. v. L. performed calculations of the XMLD signal. N. W. P. performed the micromagnetic simulations. P. G. R conceived and designed the experiment and supervised the analysis together with R. D. J, while C.-B. E. supervised the film growth. M. S. R. supervised the MOKE measurement. P. G. R. and F. P. C prepared the first draft of the manuscript. All authors discussed and contributed to the manuscript.

\newpage


\begin{figure}[!h]
\captionsetup{labelformat=empty}
\centering
\includegraphics{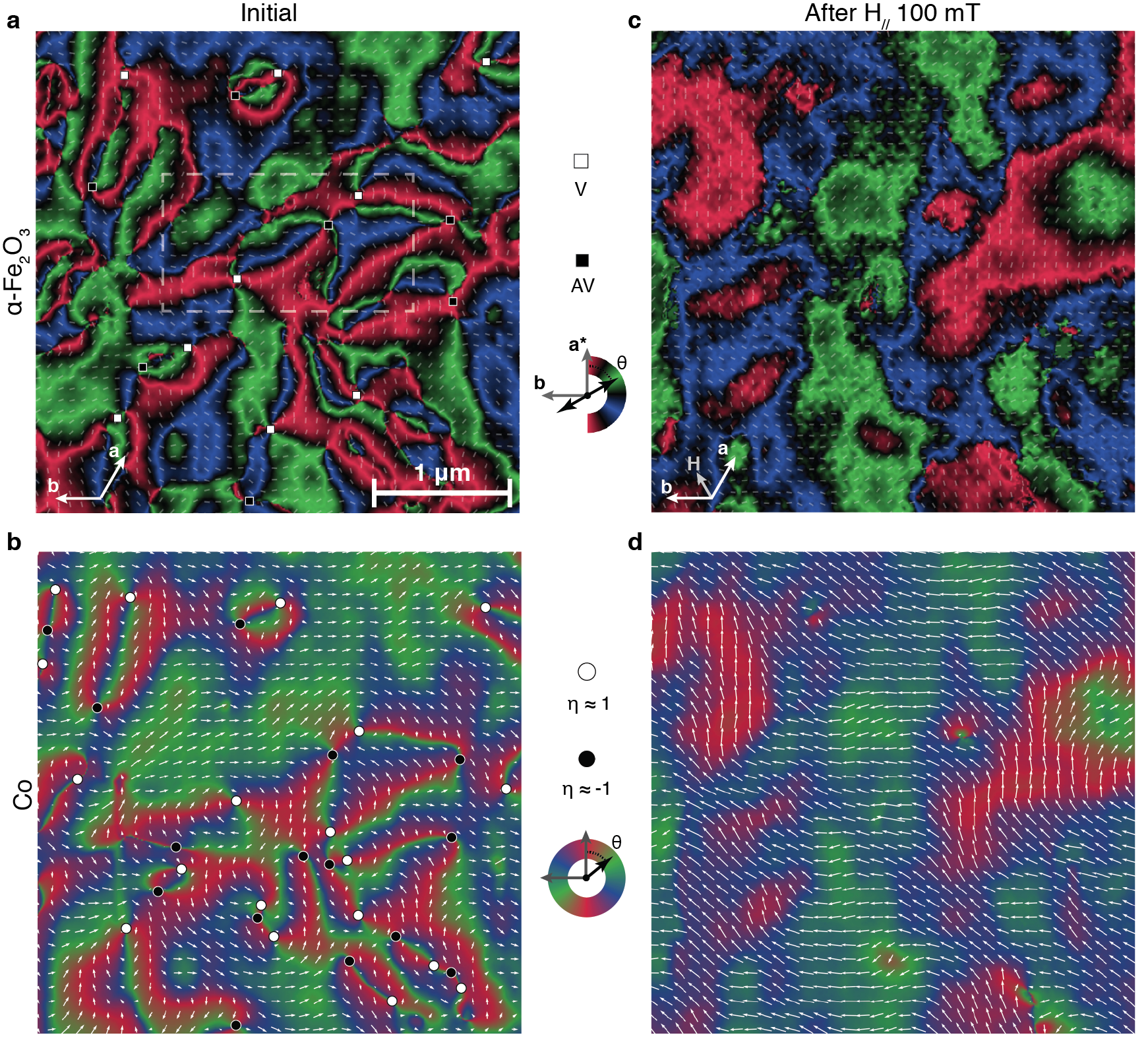}
\caption{}
\end{figure}
\clearpage
\begin{figure}
    \captionsetup{labelformat=adja-page}
    \ContinuedFloat
    \caption{\textbf{Magnetic domain structure of \hematite{} / Co film. a,} Initial \hematite{} antiferromagnetic domain structure as determined by vector - mapped XMLD-PEEM. White (black) squares indicate the approximate centre of identified vortex (anti-vortex) cores. White bars display the measured antiferromagnetic spin direction for a ten square pixel region; color represents the angle of the antiferromagnetic spin direction for each pixel measured clockwise from a$^*$. Black regions highlight AFM spin directions deviating significantly from the three expected AFM spin directions of \hematite{}, primarily highlighting domain walls. The dashed white box shows the region of interest used in \cref{fig:fe2o3_co_comparison}. \textbf{b,} Co spin vector map determined from XMCD-PEEM, white (black) circles are topological defects with a winding number of 1 (-1), see Methods for details. White arrows show the local in-plane spin orientation of the Co film for a ten square pixel region and color the angle of the spin, for each pixel, measured clockwise from a$^*$. \textbf{c-d,} \hematite{}  and Co vector map after application of an \textit{ex-situ}, 100 \si{\milli\tesla} magnetic field along [1,1,0]. All images were recorded at the same position on the sample surface.}\label{fig:fe2o3_domain_maps}
\end{figure}

\begin{figure}[!h]
\centering
\includegraphics{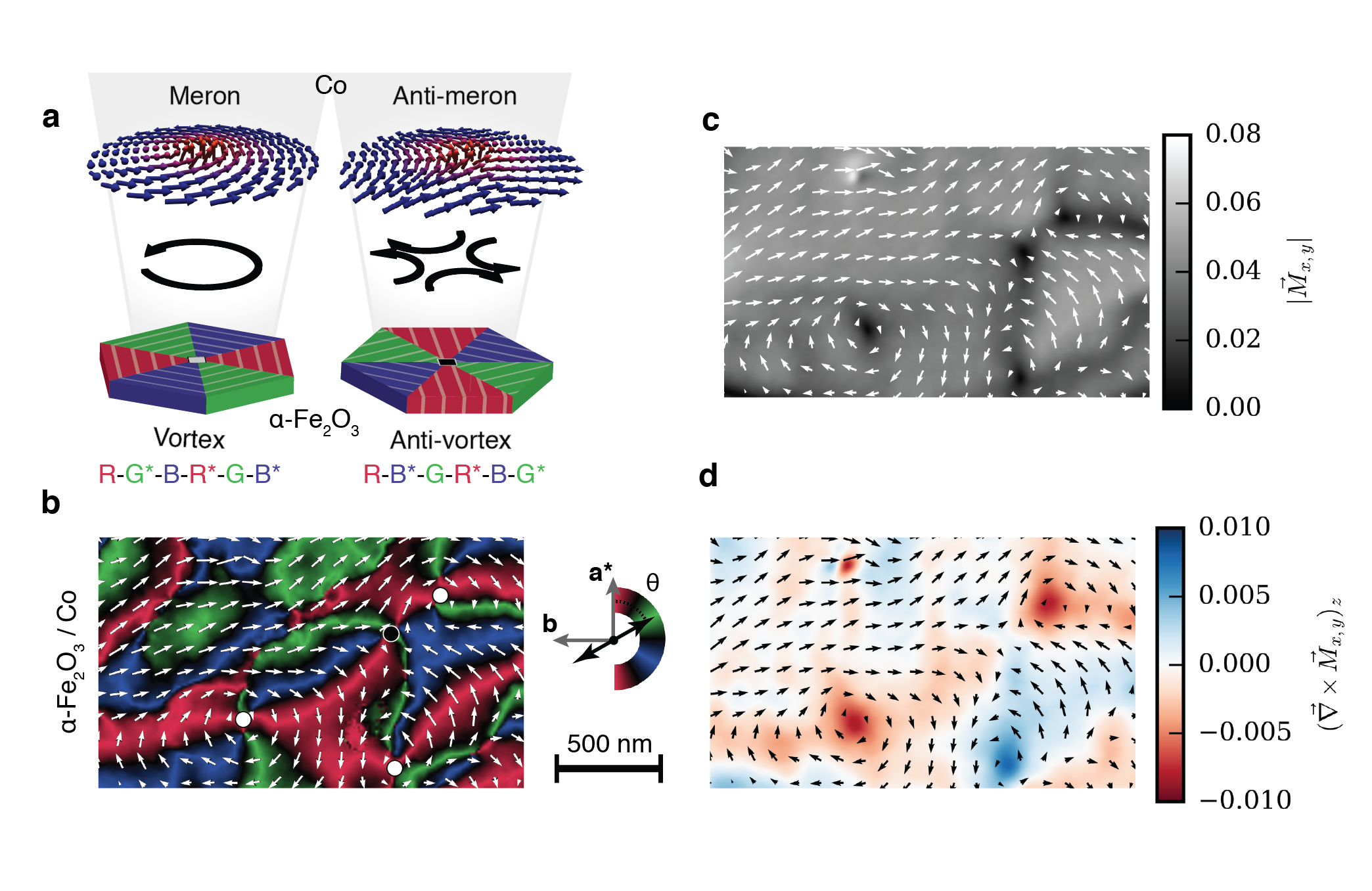}
\caption{\textbf{Topological defects in \hematite{} and Co. a,} Graphic demonstrating coupling between \hematite{} (anti-)vortices and Co (anti-)merons. \textbf{b,} Combined ferromagnetic (white arrows) and antiferromagnetic (colour) \hematite{} / Co film domain structure for a small area of our film, shown by the dashed box in \cref{fig:fe2o3_domain_maps}a. White (black) circles show topological defects with a winding number of +1(-1). \textbf{c} Measured magnitude of Co magnetisation in the sample plane. The intensity reduction at the vortex cores is compatible with the meron picture (see text), although resolution smearing could produce similar effects. \textbf{d,} $\hat{z}$ component of the curl of the Co vector map.}\label{fig:fe2o3_co_comparison}
\end{figure}

\begin{figure}[!h]
\centering
\includegraphics{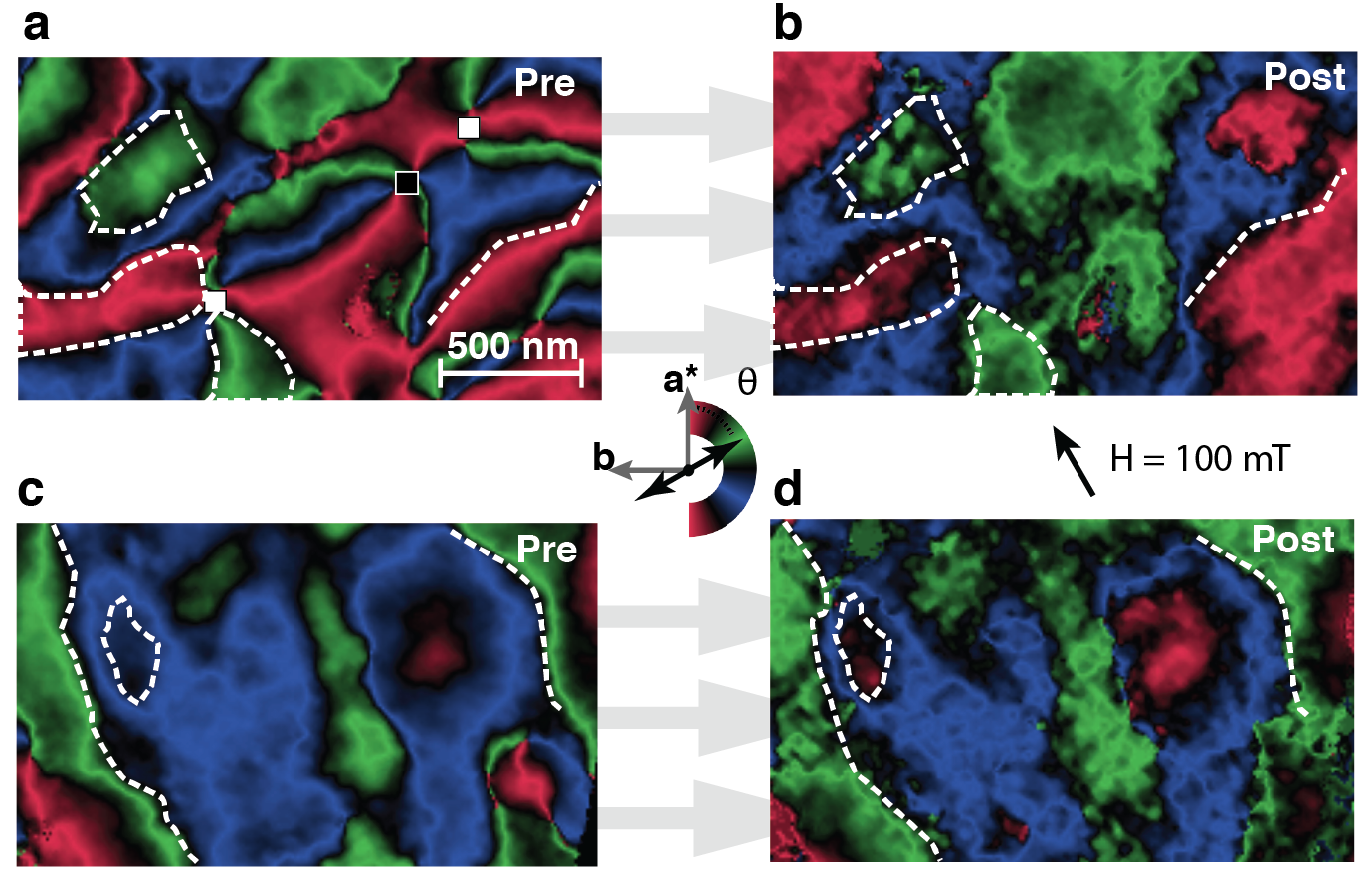}
\caption{\textbf{Vortex annihilation with \textit{ex-situ} applied field. a-d,} Measured \hematite{} domain configuration in two different regions of the sample before, a and c, and after, b and d, application of the 100 mT in-plane ([1,1,0]) magnetic field. The Co spins share a similar evolution see \cref{fig:fe2o3_domain_maps} b and d.}
\label{fig:merons}
\end{figure}

\clearpage
\putbib[fe2o3_short.bib]
\end{bibunit}
\clearpage

\end{document}